\documentclass[aps, prl, twocolumn, superscriptaddress, amsmath,  tightenlines, longbibliography]{revtex4-2}

\usepackage{tikz}

\usepackage{graphicx} 
\usepackage{epstopdf}
\usepackage{dcolumn}
\usepackage{graphicx}
\usepackage{mathrsfs}
\usepackage{subfigure}
\usepackage{booktabs}
\usepackage{amsmath}
\usepackage{physics}
\usepackage{dsfont}
\usepackage{amstext}
\usepackage{amssymb}
\usepackage{amsbsy}
\usepackage{bbm}
\usepackage{amsthm}
\usepackage{graphicx}
\usepackage{color}
\usepackage{cancel}
\usepackage{CJK}
\usepackage[colorlinks,citecolor=blue]{hyperref}

\usepackage{url}
\usepackage[colorlinks]{hyperref}
\hypersetup{%
	plainpages=true,
	breaklinks=true,       
	hypertexnames=false,  
	pageanchor=true,
	colorlinks=true,
	linkcolor={blue},
	citecolor={red},
	urlcolor={blue},
	anchorcolor={black}
}

 \makeatletter

\newcommand{\Rmnum}[1]{\expandafter\@slowromancap\romannumeral #1@}
\makeatother

\begin{document}

	\title{Finite-temperature topological invariant for higher-order topological insulators}
	\author{Congwei Lu}
	\affiliation{School of Physics and Astronomy, Applied Optics Beijing Area Major Laboratory, Beijing Normal University, Beijing 100875, China}
	\affiliation{Department of Physics, Hong Kong Baptist University, Kowloon Tong, Hong Kong, China}
	\affiliation{Key Laboratory of Multiscale Spin Physics, Ministry of Education, Beijing Normal University, Beijing 100875, China}
	\author{Lixiong Wu}
	\affiliation{School of Physics and Astronomy, Applied Optics Beijing Area Major Laboratory, Beijing Normal University, Beijing 100875, China}
	\affiliation{Key Laboratory of Multiscale Spin Physics, Ministry of Education, Beijing Normal University, Beijing 100875, China}
	\author{Qing Ai}
	\email[E-mail: ]{aiqing@bnu.edu.cn}
	\affiliation{School of Physics and Astronomy, Applied Optics Beijing Area Major Laboratory, Beijing Normal University, Beijing 100875, China}
	\affiliation{Key Laboratory of Multiscale Spin Physics, Ministry of Education, Beijing Normal University, Beijing 100875, China}


	\begin{abstract}
     We investigate the effects of temperature on the higher-order topological insulators (HOTIs). The finite-temperature topological invariants for the HOTIs can be constructed by generalizing the Resta's polarization for the ground state to the ensemble geometric phase (EGP) for the mixed states, [C.-E. Bardyn, L. Wawer, A. Altland, M. Fleischhauer, and S. Diehl, \href{https://link.aps.org/doi/10.1103/PhysRevX.8.011035}{Phys. Rev. X 8, 011035 (2018)}]. The EGP is consistent with the Resta's polarization both at zero temperature and at finite temperatures in the thermodynamic limit. {We find that the temperature can change the critical point and thus induces a phase transition from a topologically-trivial phase to a nontrivial phase in a finite-size system, manifesting changes in the winding of the EGP.}
	\end{abstract}
	
	\maketitle
	
	\textit{Introduction}.---The phases of quantum matter can be characterized and classified by the topology of their ground states \cite{Wen2017rmp,PhysRevB.55.1142,PhysRevB.78.195125,ryu2010topological}. The topological properties of the quantum phases, such as the protected edge states and the quantized particle transport, all originate from the topological invariants of the bulk state \cite{Thouless1982PRL,Thouless1983PRB3,RevModPhys.82.1959,RevModPhys.82.3045,Qi2011RMP}, which can only be defined in the ground state. Therefore, how to characterize the topology of the finite-temperature quantum states or even the non-equilibrium steady states of open quantum systems is an important issue\cite{uhlmann1986parallel,PhysRevB.88.155141,PhysRevB.91.165140,PhysRevLett.112.130401,diehl2011topology,Bardyn_2013,PhysRevA.91.042117,PhysRevA.98.052101,PhysRevLett.127.250402,PhysRevB.94.201105,PhysRevLett.131.083801,PhysRevB.103.035112,PhysRevLett.127.086801,PhysRevX.11.021037,PhysRevLett.124.040401,PRXQuantum.5.030304}. Recently, it has been shown that the many-body Resta's polarization \cite{Resta1998prl} defined in the ground state can be generalized to the phase of the expectation value of the polarization operator in the Gaussian mixed states of fermions in 1D, i.e., the ensemble geometric phase (EGP), which can be directly measured \cite{PhysRevX.8.011035}. The winding of the EGP of the mixed states upon cyclic-parameter variation yields a quantized topological invariant, which can be related to the quantized particle transport in an auxiliary system weakly	coupled to the fermion chain \cite{PhysRevA.104.012209}. The concept of the EGP can be generalized to 2D systems \cite{PhysRevB.104.094104}, the interacting systems \cite{PhysRevLett.125.215701}, the critical systems \cite{Balabanov2022PRB}, the systems with time-reversal symmetry \cite{PhysRevB.104.214107,Pi2022prb}, and the dissipative systems \cite{PhysRevResearch.5.023004,Mao_2024}. However, all these extensions are limited to the first-order topological systems.
	
	Recently, the higher-order topological insulators (HOTIs) have attracted broad interest due to their unconventional bulk-boundary correspondence \cite{ZhangF2013PRL,Benalcazar2017Science,Benalcazar2017PRB5,Langbehn2017PRL1,SongZ2017PRL2,Kunst2018PRB5,Peterson2018Nature,Serra2018Nature,Geier2018PRB5,Ezawa2018PRL1,Schindler2018SA,ZhangX2019NP,NiX2018NM,XueH2019NM,Khalaf2018PRB6,Schindler2018NP,Park2019PRL3,YangY2020PRR9,ChenR2020PRL3,ZengQ2020PRB4,Banerjee2020PRL1,LiuZ2021PRB5,HuaC2023PRB4,LiuB2021PRL1,Slager2015PRB6,Ghosh2021PRB8}. {In higher-order topological phases, a $d$-dimensional $n$th-order $(n\geq2)$ topological system hosts topologically-protected gapless states on its $(d\!\!\!-\!\!\!n)$-dimensional boundaries. Up to now, much effort has been devoted to understanding the effects of disorders \cite{WangC2020PRR1,LiC2020PRL1,YangY2021PRB8}, electron-electron
	interactions \cite{PengC2020PRB0,Kudo2019PRL2,ZhaoP2021PRL1} and electron-phonon interaction \cite{Lu2023aPRB} on the higher-order topological phase transitions.} Nevertheless, the finite-temperature effects on the higher-order topological phase transitions have not yet been studied. One may wonder whether the EGP can be generalized to the HOTIs? {Does the temperature affect the higher-order topological phase transitions?}
	
	 In this {\it Letter}, we propose a topological invariant to characterize the topology of the HOTIs at finite temperatures. And thus it enables us to study the effects of the temperature on the higher-order topological phase transition. To be specific, we introduce the many-body Resta's polarization of the non-interacting fermionic Benalcazar-Bernevig-Hughes (BBH) model \cite{Lu2023PRB1,WuY2022PRA1} at zero temperature and generalize it to the finite-temperature case, thereby constructing a higher-order EGP. We show that the higher-order EGP can restore the Resta's polarization at zero temperature, while at finite temperatures, the higher-order EGP is consistent with the Resta's polarization in the thermodynamic limit. The winding of the higher-order EGP is quantized and thus can be utilized to characterize the higher-order topological phase transition at finite temperatures. {We find that the temperature can affect the critical point of a finite-size system and induce a topological phase transition from a topologically-trivial phase to a topologically-nontrivial phase.}
	
\begin{figure}[!tb]
		\centering
		\includegraphics[width=8.4cm]{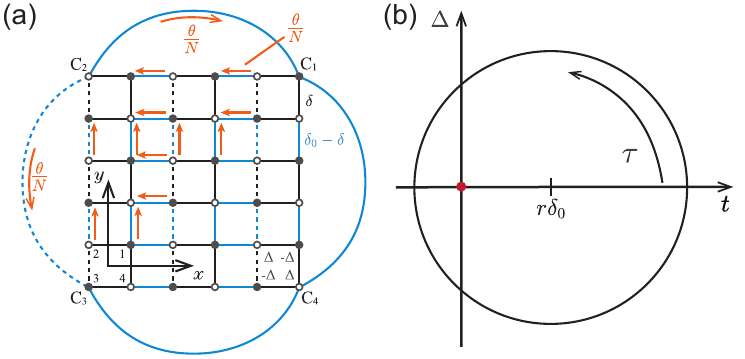}
		\caption{(a) The square lattice of BBH model with time-dependent nearest-neighbor intracell hopping constant $\pm\delta$ and intercell hopping constant $\pm(\delta_0-\delta)$, and onsite potentials {$\pm\Delta$}. Solid lines take positive signs, while dashed lines take negative signs. The BBH model under CPBC with additional flux $\theta$ in top and left supercells, and the net flux in each bulk-plaquette is $\pi$. The orange arrows indicate that the hopping along the direction of the arrow needs to be multiplied by a phase factor $e^{i\theta/N}$.  (b) The pumping circle in parameter space $(t,\Delta)$. The arrow marks the evolution direction of the pump, and the red dot represents the gapless point of system.  }\label{Fig1}
	\end{figure}

	\textit{Higher-order topological invariant for non-interacting fermionic model}.---As shown in Fig.~\ref{Fig1}(a), we consider a noninteracting fermionic BBH square-lattice model\cite{Benalcazar2017Science,Benalcazar2017PRB5} with staggered potential and periodic driving. The Hamiltonian with $N \times N$ unit cells reads
	\begin{align}\label{MIH1}
		\mathcal{H}_0 = ~&-\sum_{\langle i,j\rangle} \lambda_{ij}\hat{a}_i^\dagger \hat{a}_j
		+ \Delta\sum_{x,y=0}^{N} \sum_{\mu=1}^{4} (-1)^{\mu}\hat{n}_{x,y;\mu} ,
	\end{align}
	where $\hat{a}_i^\dagger$ is a fermionic creation operator at site $i=(x,y;\mu)$, $x,y\in\{0,1,\cdots,N-1\}$ are the positional coordinates of the unit cell, $\mu$ labels the four sites in each unit cell. $\hat{n}_{i}=\hat{a}_i^\dagger\hat{a}_i$ denotes the fermion number operator.  The nearest-neighbor-hopping energy $\lambda_{ij}$ in the unit cell and between two unit cells are respectively $\pm\delta$  and $\pm(\delta_0 - \delta)$. $\pm\Delta $ are the staggered onsite potentials. The pump cycle is parameterized by $\tau\in [0,2\pi)$, with {$t=(\delta_0-\delta)-\delta= \delta_0(\cos\tau+r)$} and $\Delta = \delta_0\sin\tau$, where $\delta=\delta_0(1-\cos\tau-r)/2$. As shown in Fig.~\ref{Fig1}(a), the solid lines take positive signs, while the dashed lines take negative signs. The undriven BBH model exhibits a second-order topological phase when $t>0$, featured by the appearance of in-gap localized states on the corners of the lattice \cite{Benalcazar2017Science,Benalcazar2017PRB5}. The pumping circle with the center localized at $(t,\Delta)=(r\delta_0, 0)$ in the parameter space $(t,\Delta)$ is shown in Fig.~\ref{Fig1}(b). When a pumping circle encloses the gapless point, i.e., $(t,\Delta)=(0,0)$, as shown in Fig.~\ref{Fig1}(b), the quantized charge will be transferred from the two anti-diagonal corners to the two diagonal corners during an adiabatic cyclic evolution in the half-filling \cite{Wienand2022PRL2}.

	Now, we review the generalization of Resta polarization to 2D hardcore Bose-Hubbard model as discussed in Ref.~\cite{Wienand2022PRL2}. And we apply it to the non-interacting fermion BBH model.	
	Following Ref.~\cite{Wienand2022PRL2}, in order to investigate the charge-transport properties on higher-order corner states, we connect the four corners of the system so that the current can pass through all four corners, and then study the total amount of charge flowing through corner $c_2$ during one pumping cycle.
	
	The total bulk Hamiltonian with corner periodic boundary conditions (CPBCs) reads $\mathcal{H}^{\text{CPBC}}=\mathcal{H}_0+\mathcal{H}^{\text{C}}$, where the corner-connecting link Hamiltonian is
	\begin{align}\label{HC}
		\mathcal{H}^{\text{C}} = (\delta-\delta_0)(\hat{a}^{\dagger}_{\text{c}_2}\hat{a}_{\text{c}_1}+\hat{a}^{\dagger}_{\text{c}_1}\hat{a}_{\text{c}_4}+\hat{a}^{\dagger}_{\text{c}_4}\hat{a}_{\text{c}_3}-\hat{a}^{\dagger}_{\text{c}_3}\hat{a}_{\text{c}_2})+\text{h.c.}
	\end{align}
	with corner sites $\text{c}_1=(N-1,N-1;1)$, $\text{c}_2=(0,N-1;2)$, $\text{c}_3=(0,0;3)$ and $\text{c}_4=(N-1,0;4)$. To define the average current operator, we first insert flux $\theta$ into two supercells, which can be described by applying the gauge transformation in corner-connecting link Hamiltonian as
	\begin{align}\label{HC_U}
		\mathcal{H}^{\text{C}}(\theta) = \mathcal{H}_0+e^{-i\theta\hat{n}_{\text{c}_2}}\mathcal{H}^{\text{C}}e^{i\theta\hat{n}_{\text{c}_2}}.
	\end{align}
The gauge transformation twists two corner-connecting links containing $\text{c}_2$, i.e., $\lambda_{\text{c}_2\text{c}_1}\to \lambda_{\text{c}_2\text{c}_1}e^{-i\theta}$ and $\lambda_{\text{c}_2\text{c}_3}\to \lambda_{\text{c}_2\text{c}_3}e^{-i\theta}$. {However, in order to apply the perturbation theory to the first order of $\tilde{\theta}=\theta/N$ when $N\to \infty$, we need to find another gauge transformation resulting in the same inserted flux $\theta$, as shown in Fig.~\ref{Fig1}(a).} It can be obtained by applying the gauge transformation
	{\begin{align}\label{V}
		\hat{V}(\tilde{\theta})=\exp(i\tilde{\theta}\hat{X})
	\end{align}} 
to the full Hamiltonian $\mathcal{H}^{\text{C}}(\theta)$ as
	{\begin{align}\label{HC_V}
		\mathcal{H}^{\text{C}}_{V}(\tilde{\theta}) = \hat{V}(\tilde{\theta})\mathcal{H}^{\text{C}}(\theta)\hat{V}^{\dagger}(\tilde{\theta}),
	\end{align}}
	where
	\begin{align}
		\hat{X}=\sum_{x\leq y}\sum_{\mu}(y-x)\hat{n}_{x,y;\mu}.
	\end{align}
	The current operator can be defined as
	\begin{align}\label{J}
		\hat{J}=-\partial_{\tilde{\theta}}\mathcal{H}^{\text{C}}_{V}(\tilde{\theta})|_{\tilde{\theta}=0},
	\end{align}
	which can be intuitively understood as the operation of transporting particles along the direction of the arrow in the top-left triangular area of Fig.~\ref{Fig1}(a). We can also define the polarization $P$ as
	{\begin{align}\label{Polization}
		P=\frac{1}{2\pi}\text{Im}~\text{ln}\left\langle\psi_0\left|\hat{V}\left(\frac{2\pi}{N}\right)\right|\psi_0\right\rangle,
	\end{align}}
	where $\ket{\psi_0}$ is the half-filling ground state of $\mathcal{H}^{\text{C}}(0)$.
	Then we can conclude that the expected value of the current operator under adiabatic evolution can be related to the derivative of $P$ with respect to the time as \cite{SM}
	\begin{align}\label{J_average}
		\frac{1}{N}\langle \hat{J}\rangle=\frac{1}{N}\text{Tr}\left[\hat{\rho}(t)\hat{J}\right]=\partial_{t}P,
	\end{align}
	where $\hat{\rho}(t)$ is the density matrix under the adiabatic evolution with Hamiltonian $\mathcal{H}^{\text{C}}(0)$.
	
	{As shown in Supplementary Material, the polarization $P$ actually corresponds to the Zak phase, i.e.,}
	\begin{align}\label{P_zak}
		P=\frac{-i}{2\pi}\int_{0}^{2\pi}d\theta\bra{\phi_0(\theta)}\partial_\theta\ket{\phi_0(\theta)}=\frac{\varphi_\text{Z}}{2\pi},
	\end{align}
	where $\ket{\phi_0(\theta)}$ is the ground state of $\mathcal{H}^{\text{C}}_{V}(\tilde{\theta})$, satisfying $\mathcal{H}^{\text{C}}_{V}(\tilde{\theta})\ket{\phi_0(\theta)}=E_0\ket{\phi_0(\theta)}$. The higher-order Zak phase $\phi_\text{Z}$ is $\mathbb{Z}_2$-quantized without the on-site potential, i.e., \cite{SM}
	\begin{align}\label{Z2}
		\varphi_\text{Z}=\pi\mathbb{Z} ~~\text{mod} ~~2\pi.
	\end{align}

	The total charge $Q_{\text{c}_2}=\int_{0}^{T}dt\langle \hat{J}\rangle/N=\int_{0}^{T}dt\partial_t P=\Delta\varphi_Z/2\pi$ flowing through corner $c_2$ during one adiabatic pumping cycle is related to the winding of the higher-order Zak phase as
	{\begin{align}\label{chern}
		Q_{\text{c}_2}&=\frac{-i}{2\pi}\int  dt\int d\theta\left[\langle\partial_t\phi_0|\partial_\theta\phi_0\rangle-\langle\partial_\theta\phi_0|\partial_t\phi_0\rangle \right].
	\end{align}}
	It is equal to the Chern number, which is gauge-invariant and integer quantized in a  cyclic system.

	\textit{Higher-order topological invariant for Gaussian mixed states}.---
	We here introduce the Gaussian mixed states of fermions, which are the mixed states of non-interacting particles. The density matrix of Gaussian states reads
	\begin{align}\label{GDM}
		\rho=\frac{1}{\mathcal{Z}}\exp\left(-\sum_{i,j}\hat{a}_i^\dagger G_{ij} \hat{a}_j\right),
	\end{align}
	with $\mathcal{Z}$ being the partition function and $\mathrm{Tr}(\rho)=1$. We consider the thermal state with
	\begin{align}\label{TDM}
		\rho&=\frac{1}{\mathcal{Z}}\exp\left(-\beta\mathcal{H}^{\text{CPBC}}\right),
	\end{align}
	{where $G=\beta \mathcal{H}^{\text{CPBC}}$ is the fictitious Hamiltonian matrix, $\beta=k_BT$ with $k_B$ and $T$ being the Boltzmann constant and temperature, respectively.}
	
\begin{figure}[!tb]
		\centering
		\includegraphics[width=8.4cm]{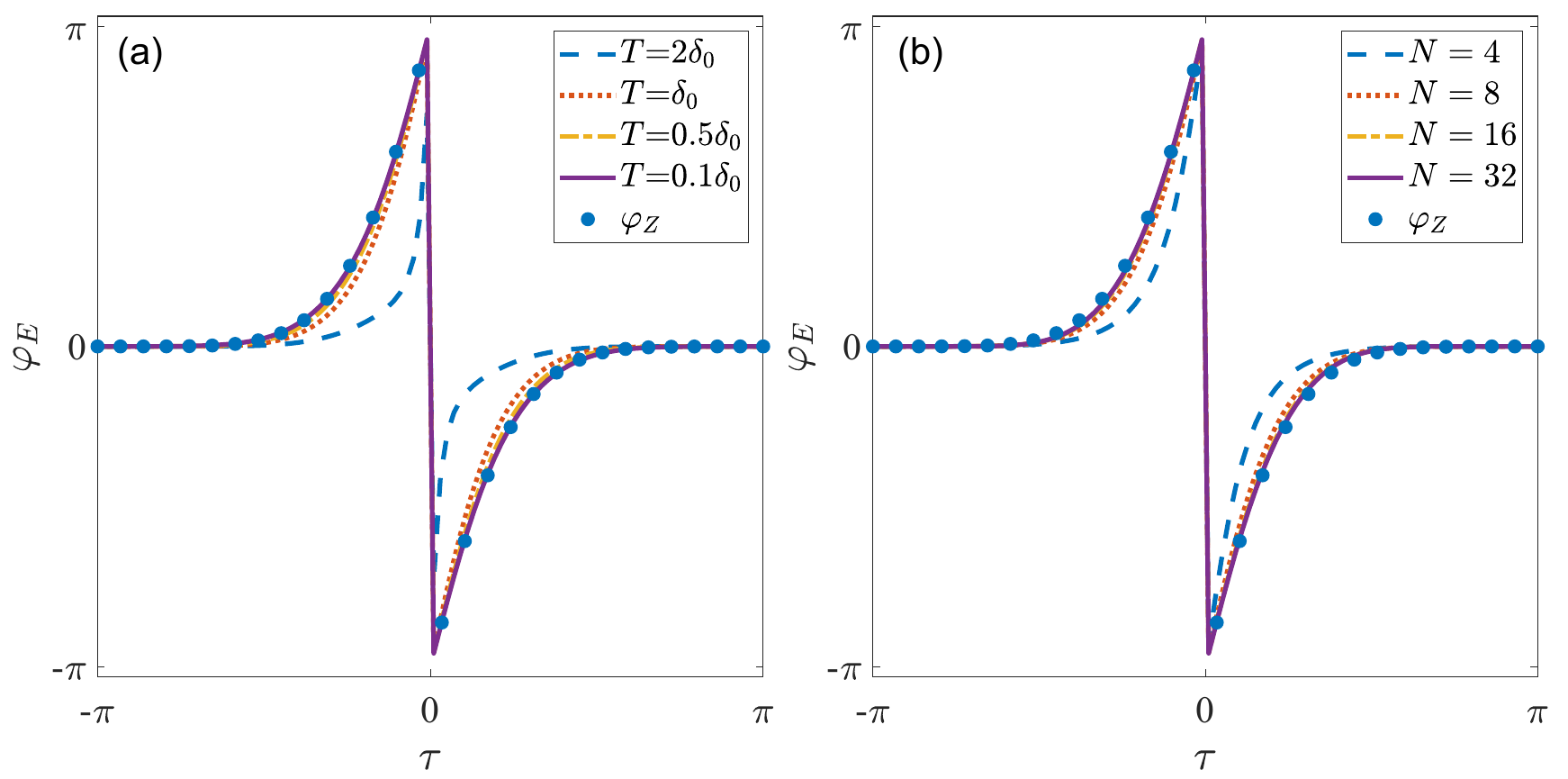}
		\caption{(a) EGP of the HOTI at different temperatures. The system includes $10\times10$ unit cells. (b) EGP of the HOTI for different sizes at temperature $T=\delta_0$. The blue dots represent the higher-order Zak phase, i.e., the EGP at zero temperature. }\label{EGP_T}
	\end{figure}

	The higher-order polarization for the ground state, defined in Eq.~(\ref{Polization}), can be generalized to the case of mixed states. Analogous to Ref.~\cite{PhysRevX.8.011035}, we can define the EGP in the HOTI as
	{\begin{align}\label{EGP}
		\varphi_E=\mathrm{Im}\ln\mathrm{Tr}\left[\rho \hat{V}\left(\frac{2\pi}{N}\right)\right]=\mathrm{Im}\ln\langle \hat{V}\rangle.
	\end{align}}
	The expectation value $\langle \hat{V}\rangle$ can be written in terms of the Gaussian integral of Grassmann number as \cite{PhysRevX.8.011035}
	\begin{align}\label{trV}
		\langle \hat{V}\rangle&=\mathrm{det}\left[-f(G)\right]\int d(\bar{\psi},\psi)e^{\bar{\psi}[f^{-1}(G)-1+V]\psi}\nonumber \\
				&=\det[1-f(G)+f(G)V].
	\end{align}
Here, $\hat{a}_i^{\dagger}$ $(\hat{a}_i)$ is replaced by the Grassmann number $\bar{\psi}$ $(\psi)$, which satisfy the Grassmann anticommutation $\psi_i\psi_j=-\psi_j\psi_i$. $V$ is the matrix representation of $\hat{V}$ and $f(G)$ is the correlation matrix with matrix elements
	 \begin{align}\label{fG}
	 	[f(G)]_{ij}=\langle\hat{a}_j^\dagger\hat{a}_i\rangle,
	 \end{align}
	 which can be analytically calculated as
	 \begin{align}\label{fG1}
	 	f(G)=\left(e^{G}+1\right)^{-1}.
	 \end{align}
	Clearly, by definition, EGP equals to the Zak phase at zero temperature. In Fig.~\ref{EGP_T}(a), we have shown the higher-order EGP $\varphi_E$ during a cycle at different temperatures of the same size. The blue solid dots represent the higher-order Zak phase defined in Eq.~(\ref{P_zak}). We can observe that, as the temperature decreases, $\varphi_E$ approaches the Zak phase $\varphi_Z$. It was shown in Ref.~\cite{PhysRevX.8.011035} that for 1D case, the EGP will reduce to the Zak phase in the thermodynamic limit, even when $T>0$~K. We discover that this phenomenon still exists in the 2D HOTI systems. Figure~\ref{EGP_T}(b) illustrates the EGP at non-zero temperatures, showing that $\varphi_E$ approaches the Zak phase $\varphi_Z$ as the size increases.
	
	\begin{figure}[!tb]
		\centering
		\includegraphics[width=8.4cm]{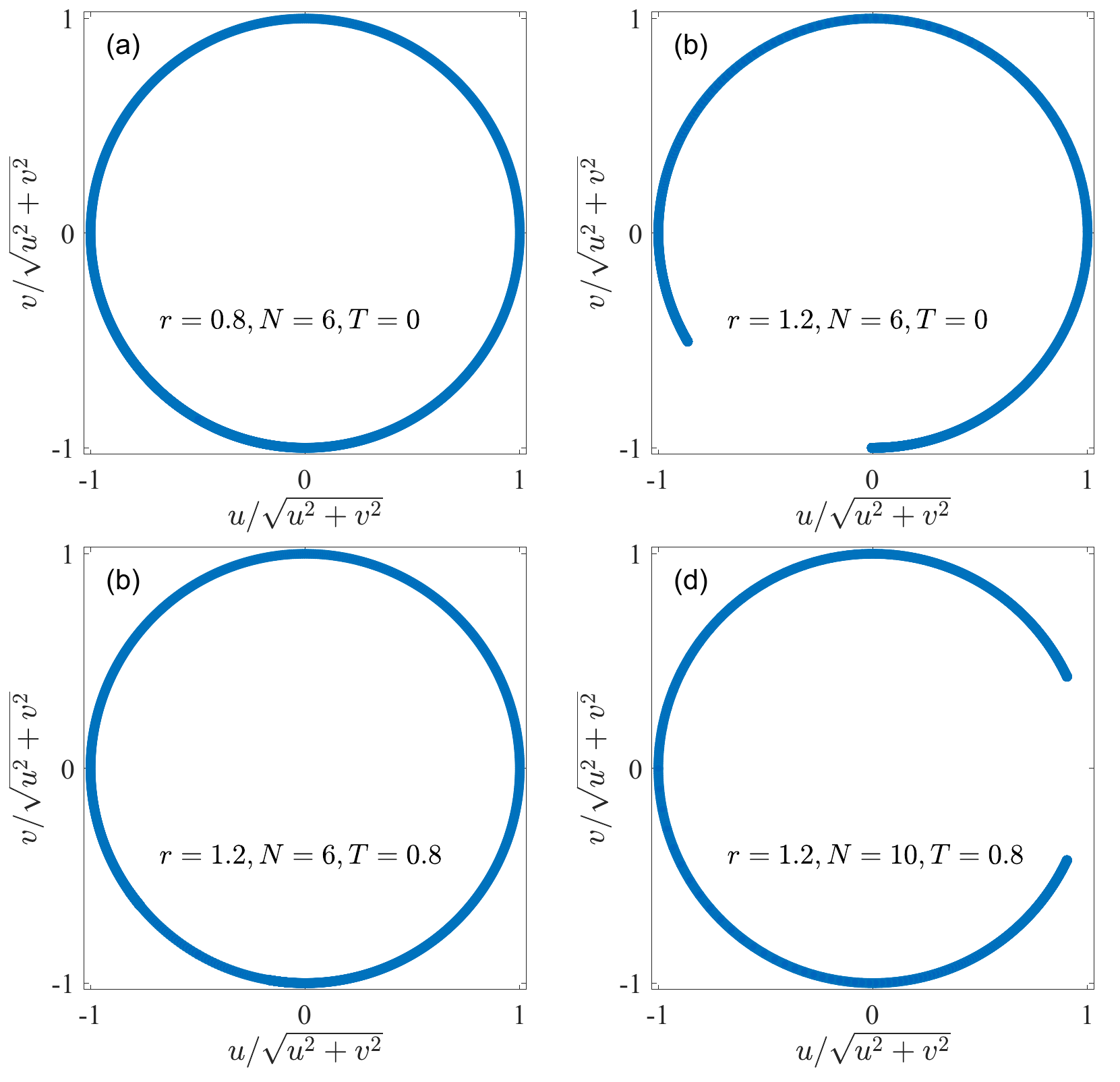}
		\caption{The trajectory of the normalized $\langle\hat{V}\rangle$ in the complex plane when $\tau$ varies from $0$ to $2\pi$. When $T=0$~K, the gapless point is (a) in (b) outside of the pumping circle. When $T>0$~K and the gapless point is outside of the pumping circle, the system size is (c) $N=6$, (d) $N=10$. }\label{EGP_N}
	\end{figure}

	We can define the higher-order topological invariant $\Delta\varphi_E$ at finite temperatures by calculating the winding of higher-order EGP upon a parameter loop \cite{PhysRevResearch.5.023004}, i.e.,
	\begin{align}\label{delta_phi_E}
		\Delta\varphi_E\!\!=\!\!\int \!\!d\tau\frac{\partial}{\partial\tau}\varphi_E\!\!=\!\!\oint_{\mathcal{P}}\left(\frac{u}{u^2+v^2}dv-\frac{v}{u^2+v^2}du\right),
	\end{align}
	where $u$ and $v$ are the real and imaginary parts of $\langle\hat{V}\rangle$, respectively. And $\mathcal{P}$ is the loop of $\langle\hat{V}\rangle$ in the complex plane. Since $\varphi_E$ is defined modulo $2\pi$, when the parameters return to their initial values after a cycle, the winding of $\varphi_E$ must be an integer multiplied by $2\pi$. It has been pointed out that the size has no contribution to the winding of $\varphi_E$, and thus $\Delta\varphi_E$ equals the winding of the Zak phase for any size $N$ \cite{PhysRevX.8.011035}. However, our results indicate that while $\Delta\varphi_E$ is always quantized for any size, the critical point of the topological phase transition sensitively depends on $N$.
	
	{$\Delta\varphi_E$ in Eq.~(\ref{delta_phi_E}) characterizes the number of times $\langle\hat{V}\rangle$ encircles the origin in the complex plane as $\tau$ varies from $0$ to $2\pi$.} In Fig.~\ref{EGP_N}, we show the trajectory of the normalized $\langle\hat{V}\rangle$. {As shown in Fig.~\ref{EGP_N}(a)-(b), at zero temperature, when the pump circle encloses the gapless point, e.g., $r=0.8$, the normalized $\langle\hat{V}\rangle$ forms a closed loop, corresponding to a nontrivial topological phase. However, when the gapless point is outside of the pump circle, e.g. $r=1.2$, $\langle\hat{V}\rangle$ can not form a closed loop, causing the system to transit to a trivial topological phase, characterized by $\Delta\varphi_E=0$. However, when the temperature increases, as shown in Fig.~\ref{EGP_N}(c), the system returns to a topologically nontrivial phase, with the normalized $\langle\hat{V}\rangle$ forming a closed loop.} Interestingly, when the size is increased, as shown in Fig.~\ref{EGP_N}(d), the normalized $\langle\hat{V}\rangle$ no longer forms a closed loop, indicating that in the thermodynamic limit, even at non-zero temperature, $\Delta\varphi_E$ remains consistent with $\Delta\varphi_Z$.
	
\begin{figure}[!tb]
		\centering
		\includegraphics[width=8.4cm]{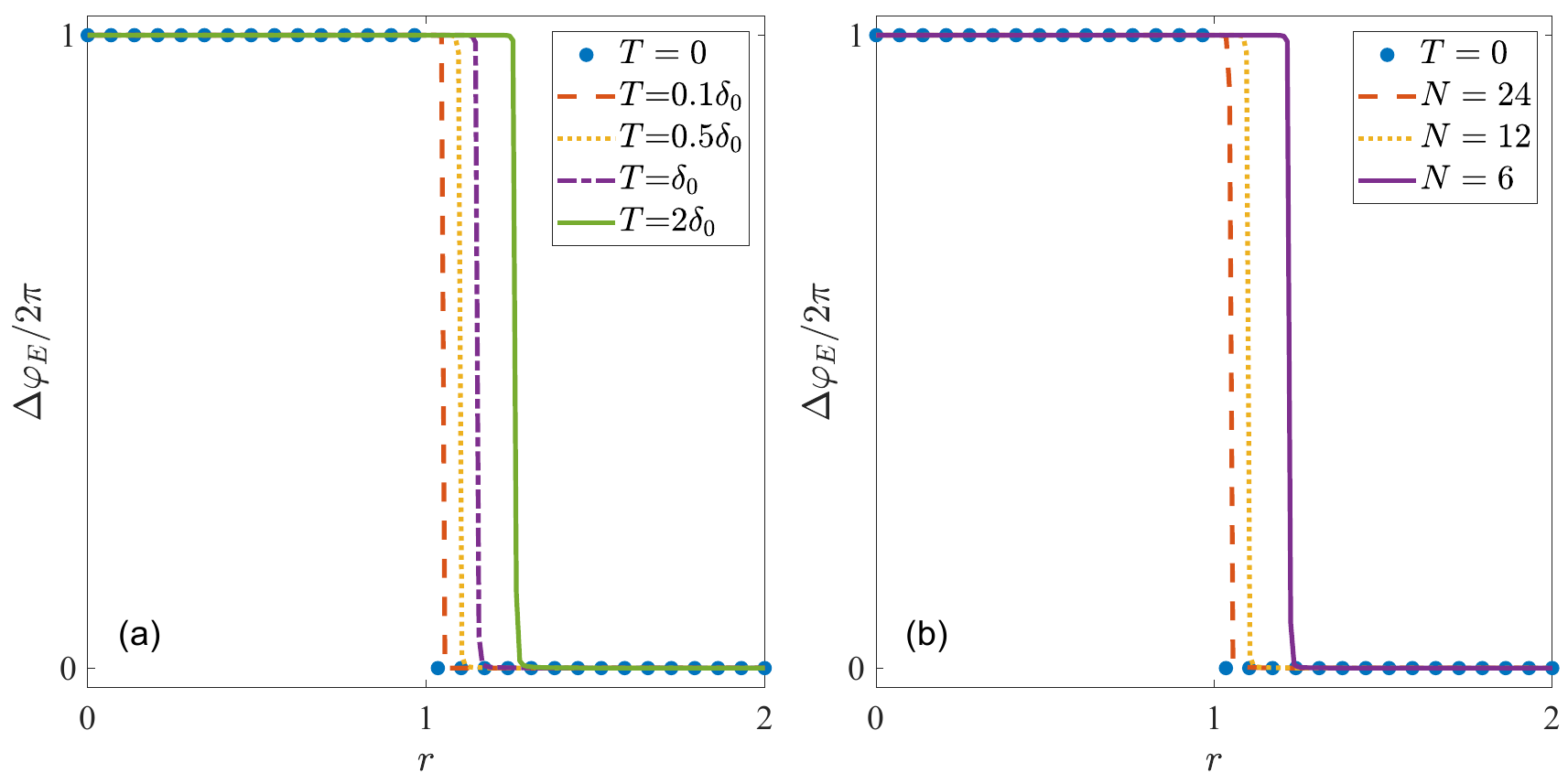}
		\caption{(a) Difference of the higher-order EGP $\Delta\varphi_E$ accumulated per cycle vs the position of the cycle center $r$ for different $N$s. The temperature of system is $T=0.5\delta_0$. (b) $\Delta\varphi_E$ vs $r$ for different $T$s. The system includes $12\times12$ unit cells. The trajectory of the parameters is shown in Fig.~\ref{Fig1}(b). }\label{chern_vs_N}
	\end{figure}

	{We show $\Delta\varphi_E$ as a function of the center of parameter circle for different sizes $N$s in Fig.~\ref{chern_vs_N}(a).} At zero temperature, the system is in a topologically-nontrivial phase when $r<1$, in which case the pumping circle encloses the gapless point. And it is in a topologically-trivial phase when $r>1$. However, when $T>0$, the position of the critical point will change. For a given temperature, as the size of the system increases, the critical point will approach the one at zero temperature. It is surprising that as the temperature increases, the critical point will be greater than 1, as shown in Fig.~\ref{chern_vs_N}(b). {It suggests that if the system is in a topologically-trivial phase near the critical point at zero temperature, increasing the temperature will induce a topological phase transition, turning the system into a topologically-nontrivial phase.}
	
\begin{figure}[!tb]
		\centering
		\includegraphics[width=8.4cm]{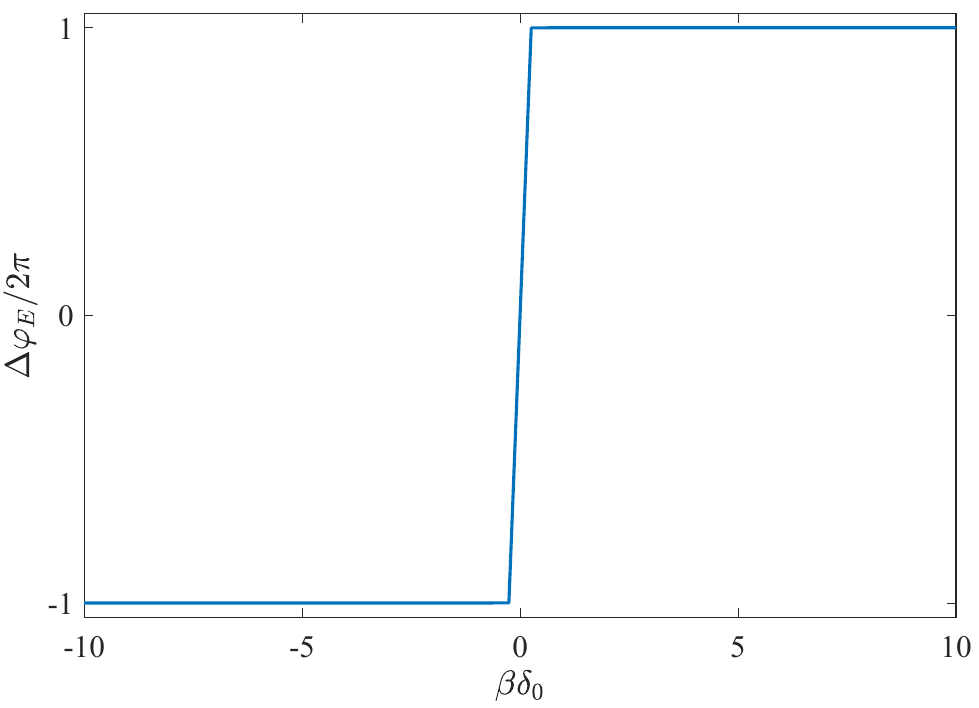}
		\caption{Difference of the higher-order EGP $\Delta\varphi_E$ accumulated per cycle encircling the gapless point vs the inverse temperature $\beta$. The trajectory of the parameters is shown in Fig.~\ref{Fig1}(b). There are $10\times10$ unit cells in the system. }\label{chern_vs_beta}
	\end{figure}

	As discussed above, the topological phase transition is dominated by the pump parameters. {On the other hand, even if the system is in a topologically-nontrivial phase and far away from the critical point, i.e., $r=1$, the change in the sign of the inverse temperature $\beta$ can also cause the closing and inversion of the bands of the fictitious Hamiltonian, thereby inducing a topological phase transition.} We demonstrate $\Delta\varphi/(2\pi)$ as a function of $\beta$ when $r=0$ in Fig.~\ref{chern_vs_beta}. When $\beta>0$, the system is in a nontrivial phase, i.e., $\Delta\varphi/(2\pi)=1$, since the parameter cycle encircles the gapless point and the bands of the fictitious Hamiltonian are gaped. {As the temperature approaches the positive infinity, alternatively $\beta\rightarrow0^{+}$, the energy gap of the fictitious Hamiltonian $G=\beta H$ vanishes.} As $\beta$ continues to decrease and even becomes negative, the energy bands of $G$ invert, leading to a topological phase transition, with $\Delta\varphi/(2\pi)$ changing from $1$ to $-1$.

	\textit{Conclusion}.---In this {\it Letter}, we introduce a topological invariant for the finite-temperature states of the HOTI. The Resta's polarization at zero temperature can be generalized to the EGP at finite temperatures in the non-interacting fermionic BBH model and the winding of the EGP yields the higher-order topological invariant $\Delta\varphi_E$. The EGP defined in higher-order topological state coincides with higher-order Zak phase in the limit $T \to 0$. In the thermodynamic limit, i.e., $N\to\infty$, the EGPs at finite temperatures coincide with the higher-order Zak phase at zero temperature. Finite-temperature higher-order topological phases are characterized by $\Delta\varphi_E$. {Our numerical simulations confirm that at finite sizes, the critical point of the topological phase transition characterized by $\Delta\varphi_E$ differs from the counterpart at zero temperature, which indicates that the temperature can induce the system to transit from a topologically-trivial phase to a topologically-nontrivial phase. But in the thermodynamic limit the critical points at different temperatures are the same.}

\textit{Acknowledgments}.---This work is supported by Innovation Program for Quantum Science and Technology under Grant No.~2023ZD0300200, Beijing Natural Science Foundation under Grant No.~1202017 and the National Natural Science Foundation of China under Grant Nos.~11674033, 11505007, and Beijing Normal University under Grant No.~2022129.
\providecommand{\noopsort}[1]{}\providecommand{\singleletter}[1]{#1}%

\end{document}